\documentclass[aps,prx,twocolumn,showpacs,superscriptaddress,amsmath,amssymb,amsfonts,floatfix,longbibliography]{revtex4-1}

\usepackage{graphicx}
\usepackage{placeins}
\usepackage{float}
\usepackage{dcolumn}
\usepackage{bm}
\usepackage{amssymb}
\usepackage[version=4]{mhchem}
\usepackage{microtype}
\newcolumntype{C}[1]{>{\centering\arraybackslash}p{#1}}
\usepackage{xfrac}
\usepackage{gensymb}
\usepackage{xcolor}
\usepackage{multirow}
\usepackage{comment}
\usepackage{times}
\usepackage[varg]{txfonts}
\usepackage{mathrsfs}
\usepackage{amsmath,amssymb,bm,mathtools}
\usepackage{amsfonts}
\usepackage{booktabs}
\usepackage{natbib,hyperref}
\hypersetup{
	colorlinks=true, 
	linkcolor=blue,  
}
\usepackage{soul}
\usepackage{epstopdf}
\usepackage{array}
\usepackage{multirow}
\usepackage{tcolorbox}
\usepackage[percent]{overpic}
\usepackage{bbm}

\usepackage{widetable}
\usepackage{array}
\usepackage{xcolor,colortbl}
\usepackage{lipsum}

\begin{document}

\title{Magnetic Field Tuning of Parallel Spin Stripe Order and Fluctuations \\ near the Pseudogap Quantum Critical Point in La$_{1.36}$Nd$_{0.4}$Sr$_{0.24}$CuO$_4$}

\author{Qianli Ma}
\affiliation{Department of Physics and Astronomy, McMaster University, Hamilton, Ontario, L8S 4M1, Canada}

\author{Evan M. Smith}
\affiliation{Department of Physics and Astronomy, McMaster University, Hamilton, Ontario, L8S 4M1, Canada}

\author{Zachary~W. Cronkwright}
\affiliation{Department of Physics and Astronomy, McMaster University, Hamilton, Ontario, L8S 4M1, Canada}

\author{Mirela Dragomir}
\affiliation{Brockhouse Institute for Materials Research, Hamilton, Ontario, L8S 4M1, Canada}
\affiliation{Electronic Ceramics Department, Jožef Stefan Institute, 1000 Ljubljana, Slovenia}

\author{Gabrielle Mitchell}
\affiliation{Department of Physics and Astronomy, McMaster University, Hamilton, Ontario, L8S 4M1, Canada}

\author{Barry~W. Winn}
\affiliation{Neutron Scattering Division, Oak Ridge National Laboratory, Oak Ridge, TN 37830, United States}

\author{Travis~J. Williams}
\affiliation{Neutron Scattering Division, Oak Ridge National Laboratory, Oak Ridge, TN 37830, United States}

\author{Bruce~D.~Gaulin}
\affiliation{Department of Physics and Astronomy, McMaster University, Hamilton, Ontario, L8S 4M1, Canada}
\affiliation{Brockhouse Institute for Materials Research, Hamilton, Ontario, L8S 4M1, Canada}
\affiliation{Canadian Institute for Advanced Research, MaRS Centre, West Tower 661 University Ave., Suite 505, Toronto, ON, M5G 1M1, Canada}

\date{\today}


\begin{abstract}

A quantum critical point in the single layer, hole-doped cuprate system La$_{1.6-x}$Nd$_{0.4}$Sr$_x$CuO$_4$ (Nd-LSCO), near $x$ = 0.23 has been proposed as an organizing principle for understanding high temperature superconductivity.  Our earlier neutron diffraction work on Nd-LSCO at optimal and high doping revealed static parallel spin stripes to exist out to the QCP and slightly beyond, at $x$ = 0.24 and 0.26.  We examine more closely the parallel spin stripe order parameter in Nd-LSCO in both zero magnetic field and fields up to 8 T for H // c in these single crystals.  In contrast to earlier studies at lower doping, we observe that H //c in excess of $\sim$ 2.5 T eliminates the incommensurate quasi-Bragg peaks associated with parallel spin stripes.  But this elastic scattering is not destroyed by the field; rather it is transferred to commensurate {\textbf{Q} = 0} Bragg positions, implying that the spins participating in the spin stripes have been polarized.  Inelastic neutron scattering measurements at high fields show an increase in the low energy, parallel spin stripe fluctuations and evidence for a spin gap, $\Delta_{spin}$= 3 $\pm$ 0.5 meV for Nd-LSCO with $x$ = 0.24.  This is shown to be consistent with spin gap measurements as a function of superconducting T$_C$ over five different families of cuprate superconductors, which follow the approximate linear relation, $\Delta_{spin}$ = 3.5 k$_B$T$_C$.

\end{abstract}

\maketitle

\section{Introduction}

Quasi-two dimensional (2D) high temperature superconducting cuprates have been of intense interest since their discovery some 35 years ago\cite{discovery1986,bkg2lbco,bkg1,theoryofhightc,bourne1987complete,kotliar1988superexchange,schrieffer1988spin,zhang1988effective,franck1994experimental,emery,kulic2000interplay,patricklee,mukuda2011high,scalapino,taillefer2010scattering,proust2019remarkable}.  A central dimension to their behaviour has been the relationship between antiferromagnetism and superconductivity and how this evolves with hole doping, as occurs when Sr$^{2+}$ is substituted for La$^{3+}$ in either La$_{2-x}$Sr$_x$CuO$_4$ (LSCO) or La$_{1.6-x}$Nd$_{0.4}$Sr$_x$CuO$_4$ (Nd-LSCO), or when with Ba$^{2+}$ is substituted for La$^{3+}$ in La$_{2-x}$Ba$_x$CuO$_4$ (LBCO)\cite{tranquada2013spins,tranquadareview2020}.  The undoped parent compound of this single layer, 214 family of cuprate superconductors is La$_2$CuO$_4$, a Mott insulator which displays commensurate, three dimensional (3D) antiferromagnetic (AF) order at $\sim$ 300 K\cite{Kastner,keimer1992magnetic}.  Doping with mobile holes quickly destroys the 3D AF order, at $x$ $\sim$ 0.2, replacing it with quasi-2D incommensurate order, that can be described within the stripe model\cite{tranquada1995evidence,tranquadareview2020}.  Superconductivity first appears at $x$ = 0.05, but the superconducting phase diagram is structured as a function of hole-doping, $x$, with a depression in T$_C$ at the ``1/8 anomaly", $x$ = 0.125, and superconducting ground states extend out to at least $x \sim$ 0.27\cite{michon2019thermodynamic,mirelakyle}.

Describing the 2D CuO$_2$ layers as edge-sharing squares for which the tetragonal $a$ = $b$ =3.88 $\AA$ lattice constants are Cu-Cu near neighbour distances,  the AF structure in the absence of holes is a ``$\pi$, $\pi$" Neel state with commensurate magnetic Bragg peaks at the ($\frac{1}{2}$, $\frac{1}{2}$, 0) positions in reciprocal space. Within the stripe model, mobile holes are accommodated inhomogeneously, with the holes being organized into quasi 1D charge stripes, separating regions that locally resemble the ``$\pi$, $\pi$" Neel state .  However the charge stripes introduce a $\pi$ phase shift between local Neel states on either side of them, resulting in a 2D incommensurate magnetic structure. As such the stripe model interleaves incommensurate charge order with incommensurate spin order, and the $\pi$ phase shift between local ``$\pi$, $\pi$" Neel states across the charge stripes implies that the incommensuration of the charge stripes is double that of the spin stripes \cite{inhomogenious,LSCOincommen,tranquada1995evidence,tranquadareview2020,stevenlandau,steven3,steven2}.

A theory framework for understanding the commensurability between spin and charge stripes was introduced some time ago \cite{stevenlandau}.  It is worth noting, however, that this is only observed at low temperatures and in the 214 single layer family of cuprate superconductors.  In other families of high T$_C$ cuprates, such stripe ordered states are not commmensurate with each other and even appear to inhabit different doping regimes, suggesting competition.  A theoretical discussion of this phenomenology has also been presented \cite{steven3}.

Static spin stripe order was first observed by neutron scattering techniques in the Nd-LSCO system at the ``1/8 anomaly", $x$ = 0.125\cite{tranquada1995evidence}.  The quasi-Bragg peaks so observed, were not resolution limited, but displayed four incommensurate magnetic peaks split off from the ($\frac{1}{2}$, $\frac{1}{2}$, 0), or ($\pi$, $\pi$, 0), position in reciprocal space\cite{tranquada1995evidence}.  Two spin stripe structures are observed as a function of hole doping.  At low hole-doping, 0.02 $\le$ $x$ $\le$ 0.05, ``diagonal spin stripes" are observed and are characterized by quasi-1D charge stripes running along (1,1,0) directions in tetragonal reciprocal space, or next-near neighbour Cu-Cu directions within the basal plane in real space\cite{diagonalstripe}. A 45 $\degree$ rotation of the spin stripe structure was observed at $x\sim$ 0.05, to a ``parallel spin stripe" structure, remarkably coincident with the transition to superconducting ground states\cite{diagonalstripe}. This was first observed in the LSCO family, and was later observed in both LBCO and Nd-LSCO\cite{diagonalstripe,dunsigerlbco,dunsigerlbco2,Kyle}.  The parallel spin stripe structure displays four quasi-Bragg peaks split off from the ($\pi$, $\pi$, 0) position in reciprocal space, at wavevectors ($\dfrac{1}{2}$,$\dfrac{1}{2} \pm \delta$, 0) and ($\dfrac{1}{2} \pm \delta$,$\dfrac{1}{2}$, 0).  The incommensuration, $\delta$, obeys the so-called Yamada law at low doping, with $\delta \sim x$, before leveling off beyond the ``1/8 anomaly", $x \sim$ 0.125 \cite{yamadarelation}.

\begin {figure}[tb]
\linespread{1}
\par
\hspace*{-0.2in}\vspace*{-0.2in}\includegraphics[width=3.5in]{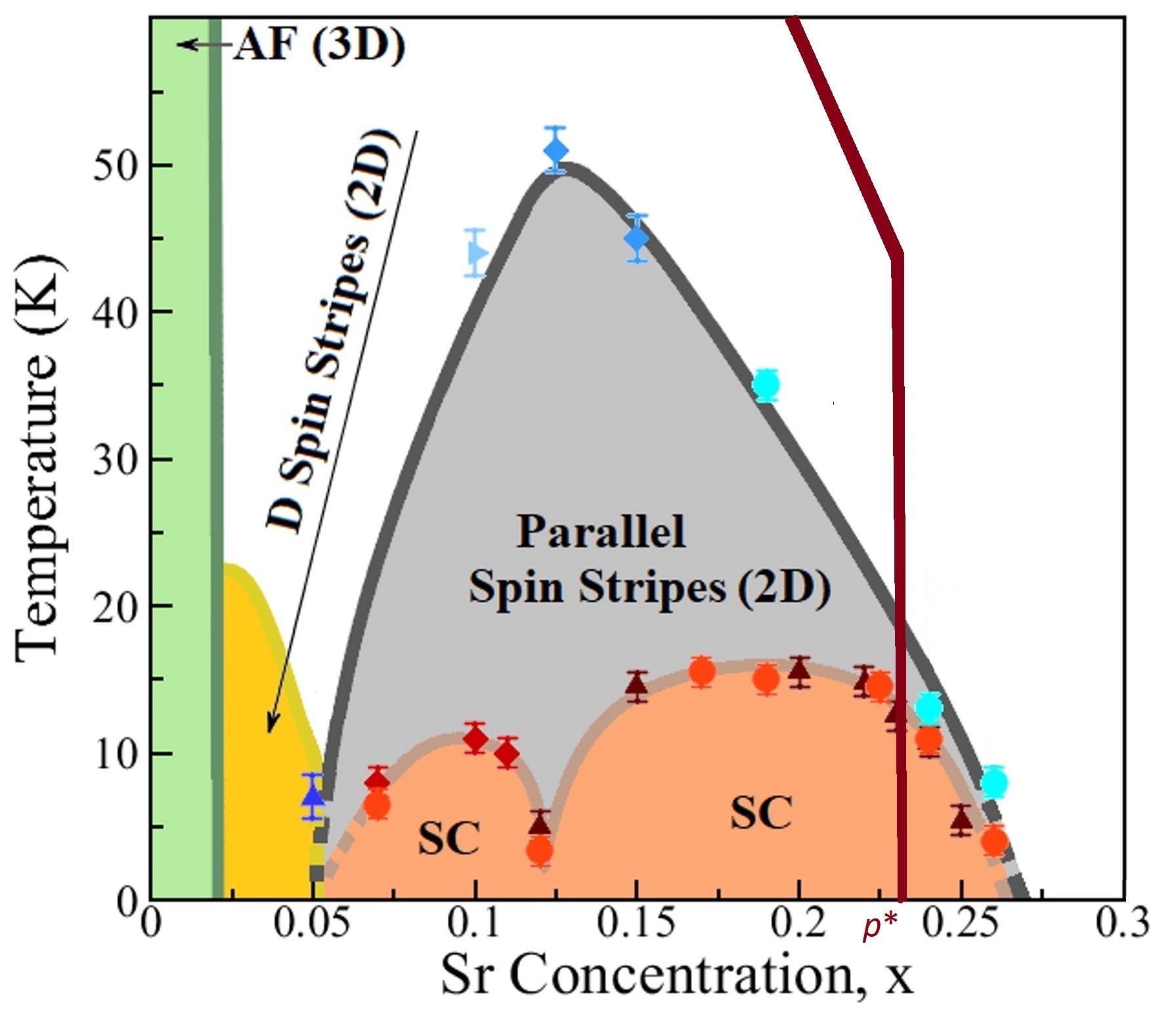}
\par
\caption{The magnetic and superconducting phase diagram for Nd-LSCO is shown, along with the line of T$^*$ vs $x$ for the pseudogap phase, which terminates at a quantum critical point, identified in \cite{michon2019thermodynamic} as $x^*$ = 0.23.  The sample under study in this paper, $x$ = 0.24, is immediately to the overdoped side of $x^*$. This plot is modified from Ref \cite{Kyle}, where the references for the magnetic and superconducting transition temperatures are given.}
\label{phasediagram}
\end{figure}

The Nd-LSCO family has gained much recent attention, primarily because its phase diagram displays the full complexity of cuprate high temperature superconductivity, while also displaying relatively low superconducting T$_C$'s (maximum T$_C$ $\sim$ 15 K), compared to the LSCO and LBCO families (maximum T$_C$'s $\sim$ 40 K and 35 K, respectively)\cite{CYRCHOINIERE2010S12,daou2009linear,badoux2016change,michon2019thermodynamic}.  Thus, experimentally practical magnetic fields can suppress superconductivity completely in the Nd-LSCO family, and its normal state properties at optimal and high doping levels can be experimentally explored. 

This has proved very useful for exploring the so-called pseudogap phase of Nd-LSCO.  Transport and thermodynamic measurements in high magnetic fields have mapped out the boundary of the pseudogap phase in Nd-LSCO \cite{michon2019thermodynamic}.  This boundary terminates at p$^*$ and is superposed in red on Nd-LSCO's magnetic and superconducting phase diagram in Fig. \ref{phasediagram}.  In particular, strong thermodynamic evidence for a pseudogap quantum critical point (QCP) has been reported in Nd-LSCO near $x$ $\sim$ 0.23\cite{daou2009linear,michon2019thermodynamic}.  This concentration is also coincident with the closing of an observed anti-nodal gap in ARPES measurements on Nd-LSCO at $x$ $\sim$ 0.24 \cite{CEMatt,arpes}. Furthermore, measurements of the Hall number, $n_H$, at low temperatures, show a transition from $n_H$ $\sim$ $x$ to $n_H$ $\sim$ 1 + $x$ at $x$ $\sim$ 0.23. While there is a compelling case for the existence of a pseudogap QCP in Nd-LSCO near $x$ $\sim$0.23, the question remains as to whether or not the pseudogap QCP is a key organizing principle in understanding cuprate superconductivity.  

Neutron scattering from hole-doped single layer cuprates in general, and Nd-LSCO in particular have been limited at optimal and high doping, due to difficulties in growing sufficiently large single crystals to allow such studies. Hence information on stripe correlations near and through the pseudogap QCP have themselves been limited.  However, recent progress has allowed a set of single crystals appropriate for neutron scattering to be grown \cite{mirelakyle}.  Combined with advances in neutron scattering instrumentation, this has allowed the recent study of both the static (on the time scale of the neutron) and dynamic parallel spin stripes in Nd-LSCO at $x$ = 0.19, 0.24, and 0.26, which spans doping levels from optimal doping, to the pseudogap QCP, to the approximate end of superconductivity\cite{Kyle,ma2021dynamic}.  Elastic neutron scattering reports static parallel spin stripes at low temperatures at all three concentrations, but with quasi-Bragg peaks characterized by decreasing in-plane correlation lengths and rapidly decreasing intensities, as a function of $x$\cite{Kyle}.  The temperature dependence of the static parallel spin stripe order, or the order parameters, allow good estimates for 2D T$_N$($x$) and these are plotted in Fig. \ref{phasediagram}.  Inelastic neutron scattering below $\hbar\omega$ = 35 meV on the same three single crystal samples, as well as an additional one at $x$ = 0.125, show little change in the dynamic spectral weight of parallel spin stripe fluctuations from $x$ = 0.125 to $0.24$, followed by an $\sim$ 1/3 reduction in spectral weight at $x$ = 0.26 \cite{ma2021dynamic}.  There is, therefore, no striking correlation observed between either the static or dynamic parallel spin stripe magnetism and the pseudogap QCP in Nd-LSCO. 

It is therefore important to examine the parallel spin stripe magnetism in Nd-LSCO near the pseudogap QCP in greater detail, and especially how it responds to an applied magnetic field.  With that in mind we have revisited the parallel spin stripe order parameter in our $x$ = 0.24 single crystal sample in zero magnetic field \cite{Kyle}, and performed new in-field order parameter measurements in H//c up to 8 T.  In this paper we combine these with low energy, inelastic neutron scattering measurements in H//c capable of examining the spin gap in the $x$ = 0.24 sample with T$_C$ = 11 $\pm$ 1 K.  

The magnetic field dependence of spin and charge stripe order in the 214 family of cuprates has an extensive history, and has been of great topical interest in the past.  Lake et al \cite{lake2002antiferromagnetic} first reported a strong increase in the strength of magnetic quasi-Bragg peaks due to parallel stripe order in the presence of an applied magnetic field in underdoped LSCO, specifically $x$ = 0.10.  These well-known results were discussed in the context of the SO(5) theory of superconductivity, where superconducting and magnetic order parameters could be rotated into each other \cite{ZhangSU5}.  Later in-field measurements on LSCO with $x$ = 0.105, 0.12, and 0.145 largely confirmed this result \cite{changmagnetic2008}, while measurements on Nd-LSCO for $x$=0.12 to H // c =10 T \cite{changmagnetic2008} and LBCO for $x$ = 0.125 to H //c = 7 T \cite{wenmagneticlbco}, showed either no field dependence, or a slight increase in the parallel spin stripe order parameter on application of the field, respectively.

Neutron diffraction measurements looking at the field dependence of spin and charge stripes were carried out on Nd-LSCO approaching optimal hole-doping, $x$ = 0.15 \cite{wakimotomag2003}.  In zero magnetic field, the parallel spin stripe order parameter shows an abrupt rise at temperatures below $\sim$ 5 K, due to the onset of Nd$^{3+}$ moments ordering and taking part in the spin stripe order. Measurements in an H //c = 7 T field suppress this low temperature component of the order parameter, and also result in an $\sim$ 20 $\%$ reduction in the order parameter at higher temperatures before going to zero at $\sim$ 40 K.  A magnetic field scan of the order parameter at very low temperatures, $\sim$ 0.1 K, shows the Nd$^{3+}$-induced enhancement of the order parameter to be largely eliminated by application of a H // c $\sim$ 0.7 T magnetic field.  Complementary neutron diffraction measurements of the parallel {\it charge} stripe order in Nd-LSCO with $x$ = 0.15 show no magnetic field dependence up to H // c = 4 T \cite{wakimotomag2003}.  

\section{Experimental Methods and Materials}
High quality single crystals of Nd-LSCO with \textit{x} = 0.24 were grown using the traveling solvent floating zone technique at McMaster University and the resulting single crystals weighed around 4 grams. The single crystals were produced using a four-mirror Crystal Systems Inc. halogen lamp image furnace at approximate growth speeds of 0.68 mm/hr, and growths lasted for approximately 1 week. The Sr concentration of this single crystal was determined by careful correlation of the structural phase transition temperatures with pre-existing phase diagrams, as described in \cite{mirelakyle}.  Further details regarding the materials preparation and single crystal growth of these samples, as well as determination of their stoichiometry, is reported in \cite{mirelakyle}. The single crystal sample used in this study displayed a mosaic spread of less than 0.5$\degree$, attesting to their high quality, single crystalline nature.  This is the same $x$=0.24 single crystal as was studied earlier in zero field in \cite{Kyle} and \cite{ma2021dynamic}.  As reported in Michon et al.\cite{michon2019thermodynamic}, heat capacity measurements in a magnetic field // c at low temperatures, T = 2 K, show saturation above 8 T in Eu-LSCO $x$ = 0.24 
and above 9 T in Nd-LSCO $x$ = 0.23.  Hence the bulk critical magnetic field // c for Nd-LSCO with x=0.24 at T= 2 K is close to 8 T.  

Triple axis neutron scattering measurements were conducted on the HB3 triple axis spectrometer (TAS) at the High Flux Isotope Reactor of Oak Ridge National Laboratory (ORNL) on the same sample $x$ = 0.24 single crystal. The sample was mounted in a pumped $^4$He refrigerator with a vertical magnetic magnetic field capability up to 8 T.  The single crystal sample was aligned with its $HK0$ scattering plane coincident with the horizontal plane, hence the applied magnetic field was applied // c. The TAS employed PG002 single crystals as both monochromator and analyzer with collimation of 48'-40'-40'-120. The scattered neutron energy was fixed at 14.7meV, allowing the use of pyrolitic graphite filter in the scattered beam, and the energy resolution of the TAS measurements performed was 1.1meV.


Complementary time-of-flight (TOF) neutron spectroscopic measurements were carried out on the \textit{x} = 0.24 single crystal using the direct-geometry, time-of-flight hybrid spectrometer, HYSPEC \cite{hyspec}, at the Spallation Neutron Source of ORNL.  All measurements were performed using \textit{E}$_i$=35 meV neutrons with Fermi chopper frequency at 180 Hz, which gave an energy resolution of $\sim$ 2.7 meV (FWHM) at the elastic position. Again the same $x$ = 0.24 Nd-LSCO single crystal sample was loaded in a pumped $^4$He magnet cryostat with a base temperature of 2 K, maximum magnetic field 8 T.   The $HK0$ plane of reciprocal space of the sample was coincident with the horizontal plane. The single crystal sample was rotated through 100$\degree$ about its vertical axis in increments of 1 degree during the course of any one measurement, which typically required 12 hours of counting time. 

\section{The parallel spin stripe order parameter in zero field in $La_{1.6-x}Nd_{0.4}Sr_{0.24}CuO_4$}

\begin{figure}[tbp]
\linespread{1}
\par
\includegraphics[width=3.4in]{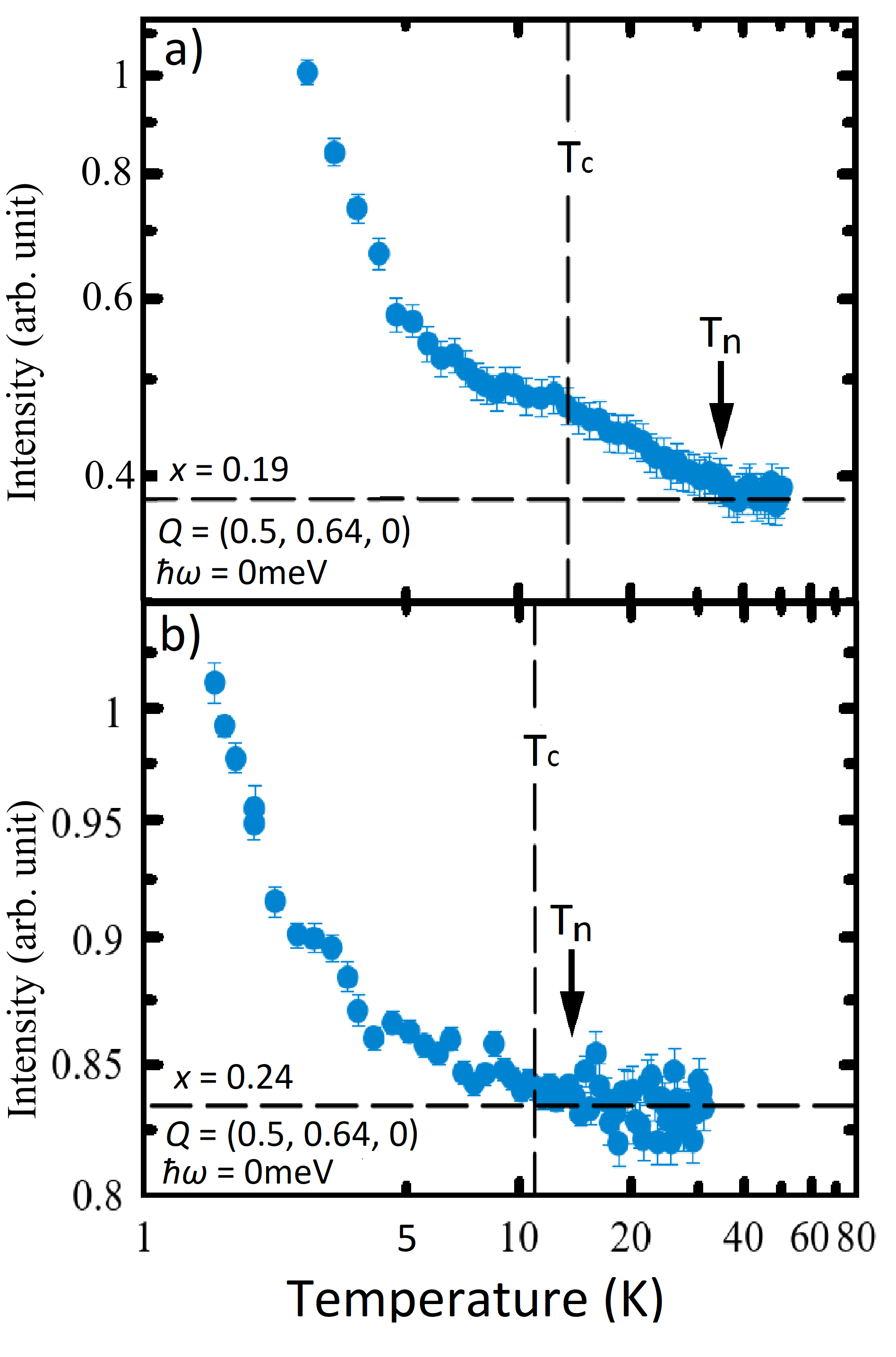}
\par
\caption{The zero magnetic field order parameters as a function of temperature for optimal doping, $x$ = 0.19, and near the pseudogap QCP, $x$ = 0.24.  Panel a) shows elastic scattering neutron data for Nd-LSCO with $x$ = 0.19, while panel b) shows the same data for $x$ = 0.24.  Fiducials are shown for both superconducting T$_C$ and 2D T$_N$ for each single crystal.  Note that both the intensity (y) scale and the temperature (x) scale are logarithmic.}
\label{OPlog} 
\end{figure}

We carried out new elastic neutron scattering measurements on the same single crystal of Nd-LSCO with $x$ = 0.24.  The new measurements were performed with a higher temperature point density and improved counting statistics, roughly four times greater than previously published data \cite{tranquadaorderparameter}.  The result is the order parameter for parallel spin stripe order in Nd-LSCO, $x$ = 0.24 shown in Fig. \ref{OPlog} b).  For comparison the corresponding order parameter for $x$=0.19, near optimal hole-doping in Nd-LSCO is shown in Fig. \ref{OPlog} a) \cite{Kyle}.  Both data sets are plotted on log-log scales.

As reported earlier, the order parameter at $x$ = 0.19 shows an onset of non-zero intensity at 2D T$_N$ = 35 $\pm$ 2 K \cite{Kyle}. The relatively large separation between 2D T$_N$ and superconducting T$_C$= 13.5 $\pm$ 1 K, allows the clear observation of an inflection point in the temperature dependence of the order parameter at T$_C$. In contrast, 2D T$_N$ = 13 $\pm$ 1 K, consistent with earlier measurements \cite{Kyle}, and superconducting T$_C$ = 11 $\pm$ 1 K \cite{mirelakyle} are almost coincident for $x$ = 0.24, as can be seen in Fig. \ref{OPlog} b).  Both order parameters show a pronounced upturn in intensity below $\sim$ 5 K, associated with the participation in ordered Nd$^{3+}$ magnetism in the parallel spin stripe structure.

\section{The magnetic field dependence of parallel spin stripe order in $La_{1.6-x}Nd_{0.4}Sr_{0.24}CuO_4$ for H // c}

\begin{figure}[tbp]
\linespread{1}
\par
\hspace*{-0.3in}\includegraphics[width=4in]{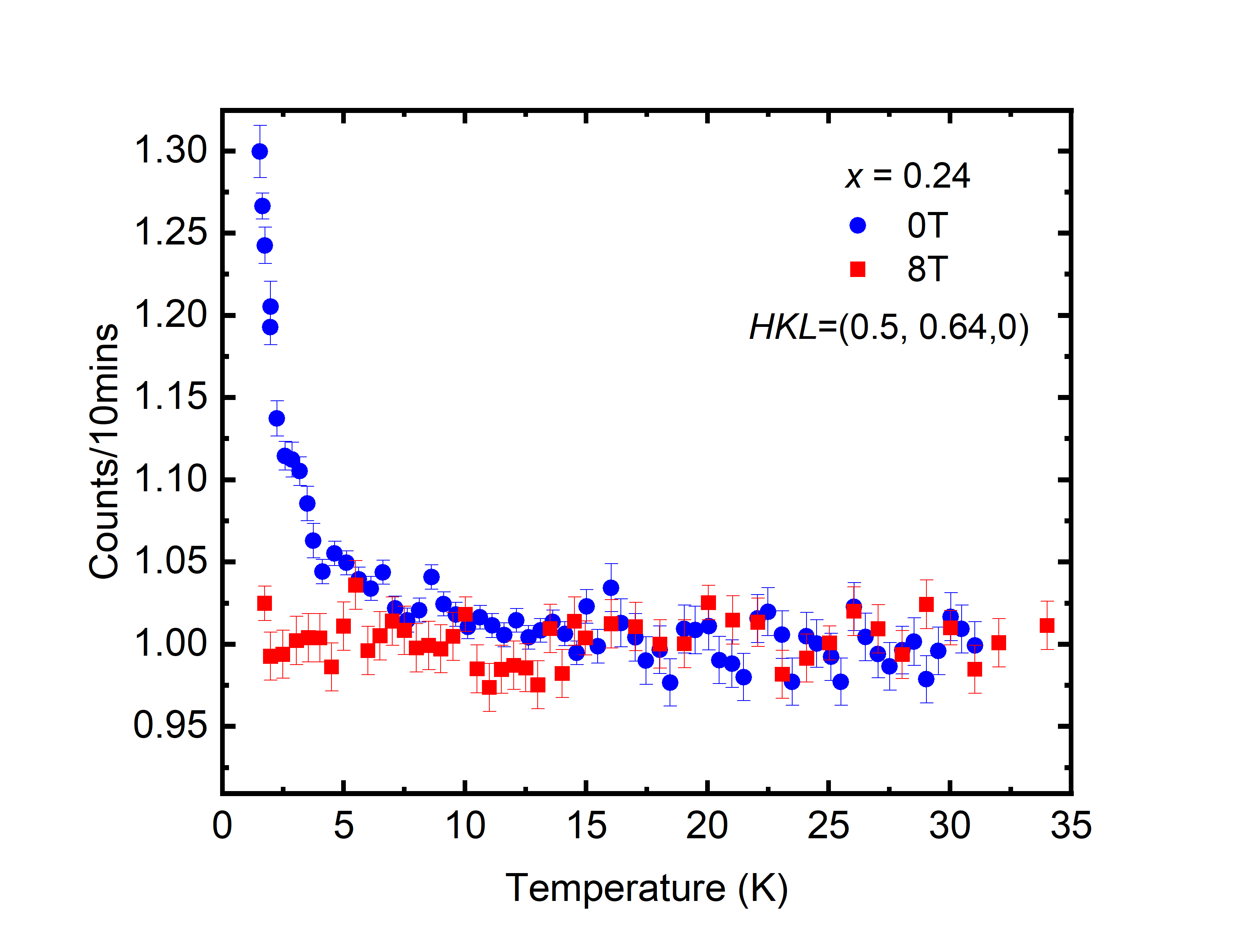}
\par
\caption{The elastic neutron scattering order parameter as measured in Nd-LSCO with $x$=0.24 is shown in both H=0 T (blue points) and H // c = 8 T (red points).  The intensity and temperature scales are linear, and the intensity has been normalized to 1 at high temperatures.}
\label{magop} 
\end{figure}

Elastic scattering order parameter measurements on our Nd-LSCO single crystal were repeated under the application of a magnetic field // c, for fields up to 8 T, and compared with those in zero field.  A subset of these measurements, at the incommensurate parallel spin stripe wavevector (0.5, 0.64, 0) with H // c = 8 T and as a function of temperature, are shown in Fig. \ref{magop}.  Here they are compared to the corresponding H = 0 measurements, also shown in Fig. \ref{OPlog} b), but now on linear scales. One can see that no temperature dependence is observed in the H // c = 8 T data set, indicating that the elastic parallel spin stripe scattering is eliminated at 8 T.

\begin{figure}[t]
\linespread{1}
\hspace*{-0.3in}
\includegraphics[width=4in]{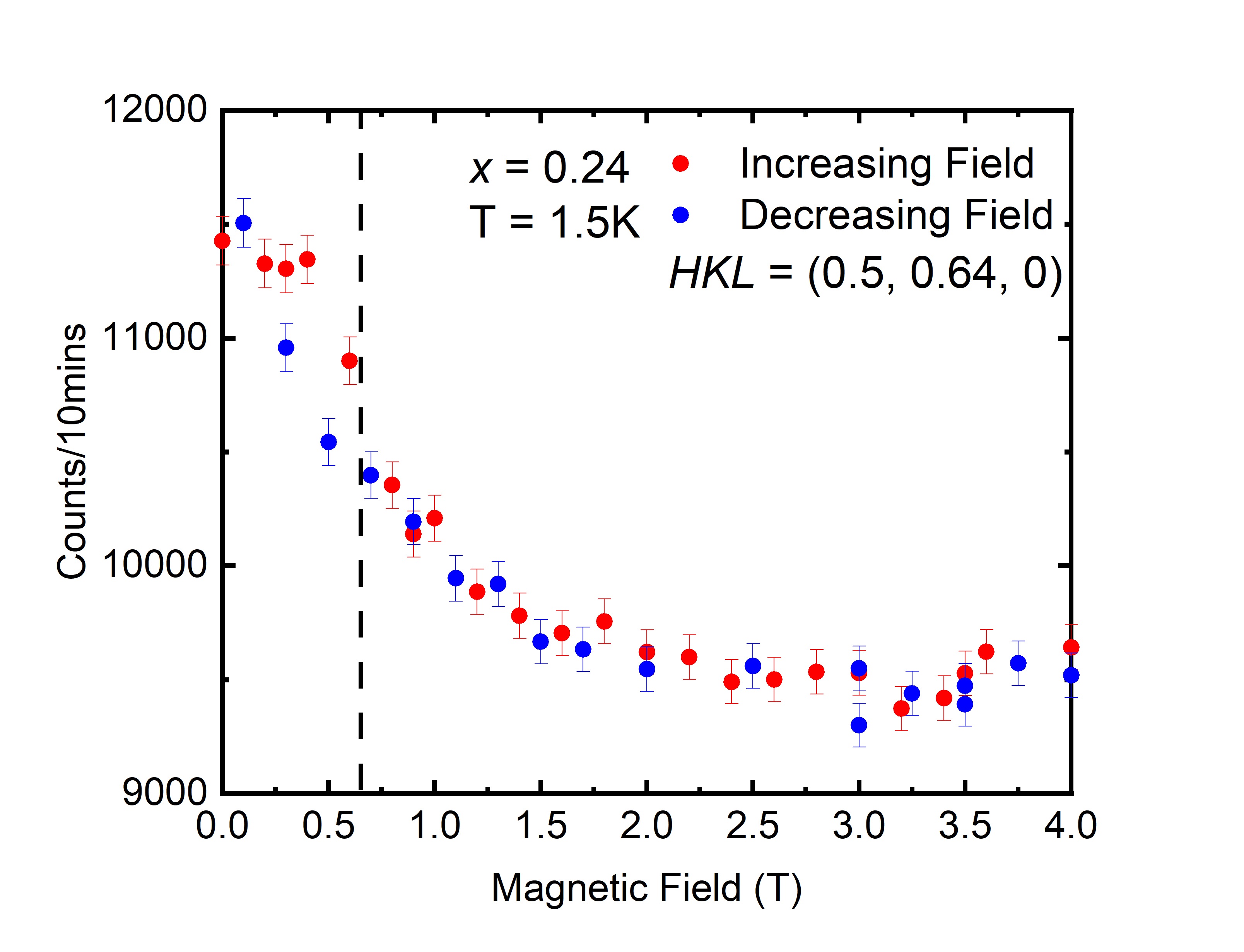}
\caption{The elastic neutron scattering parallel spin stripe order parameter as a function of H // c is shown at T = 1.5 K.  The red points show data taken while increasing H, while the blue points show data taken on decreasing H.  Note that hysteresis is present just below H = 0.7 T, which is marked as a vertical dashed line.  This indicates a 1st-order-like metamagnetic transition just below H = 0.7 T.}
\label{fieldscan} 
\end{figure}

Figure \ref{fieldscan} shows similar elastic scattering at the incommensurate parallel spin stripe wavevector, (0.5, 0.65, 0), but as a function of H // c at T = 1.5 K for the $x$ = 0.24 sample.  Measurements were performed both increasing and decreasing the magnetic field, and a clear hysteresis is observed for 0.2 T $\le$ H $\le$ 0.7 T.  We take this as evidence for a 1st order transition with field, which occurs at approximately the same field required to destroy parallel spin stripe order in Nd-LSCO $x$ = 0.15.  As can be seen from Fig. \ref{magop}, the suppression of static parallel spin stripe order in $x$ = 0.24 is complete by $\sim$ H // c = 2.5 T.

\begin{figure}[tbp]
\linespread{1}
\par
\includegraphics[width=3.5in]{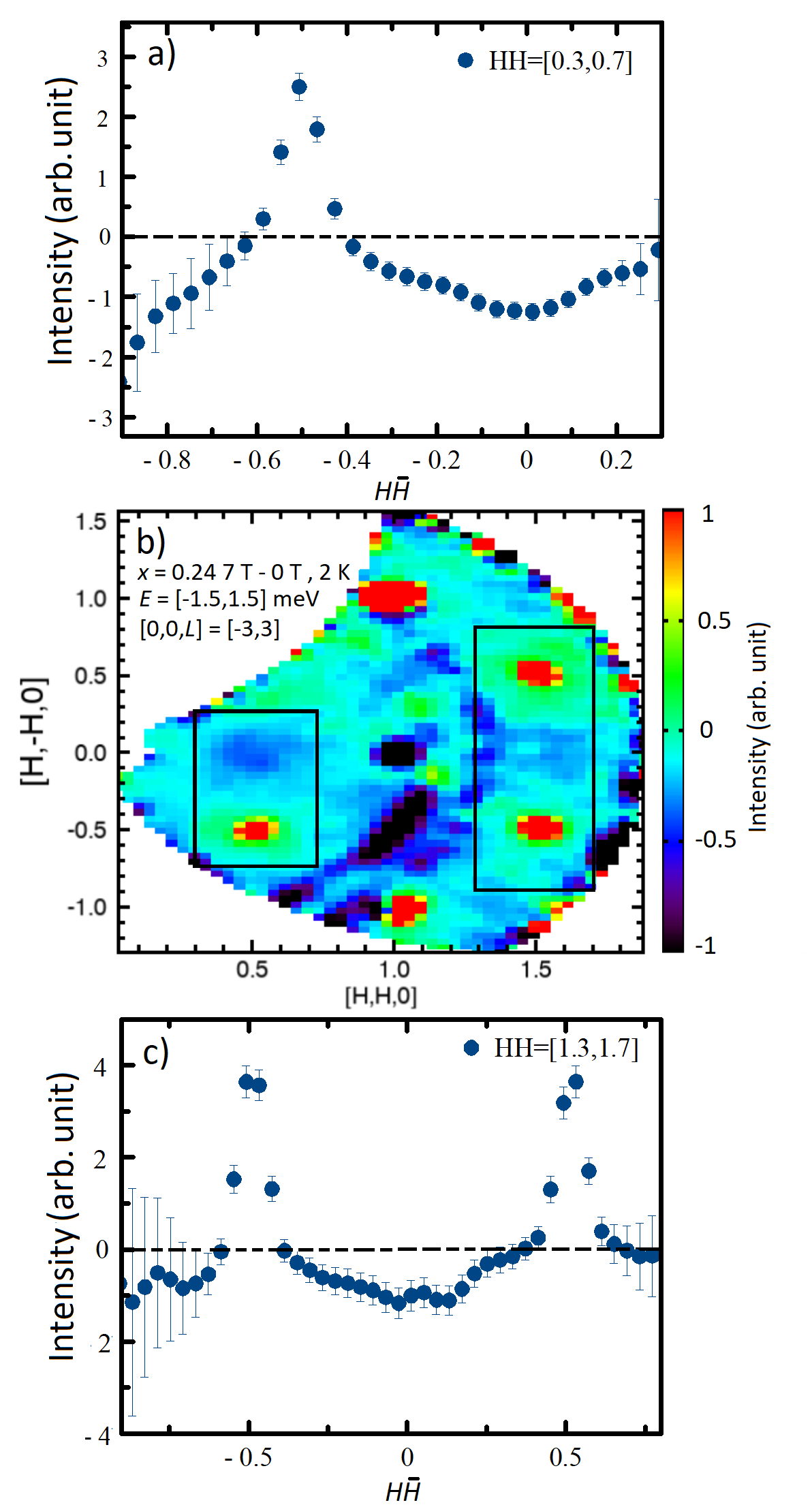}
\par
\caption{Panel b) shows a magnetic-field difference data set, H // c = 7 T - 0 T, for elastic scattering (-1.5 meV $\le$ $\hbar\omega$ $\le$ 1.5 meV) from time-of-flight neutron scattering in the [$H$,-$H$,-3 $\le$ L $\le$ 3] scattering plane.  A semi-circular region with high background from Al-powder rings due to sample addenda, which do not subtract out cleanly from each other, is present in the middle of the plot.  Panel a) shows a vertical cut through the rectangle on the left of panel b), showing a peak at the (1,0, -3 $\le$ L $\le$ 3], and a valley at (1/2 1/2 -3 $\le$ L $\le$ 3). Panel c) shows a vertical cut through the rectangle on the right of panel b), showing peaks at the (1,2, -3 $\le$ L $\le$ 3) and (2, 1, -3 $\le$ L $\le$ 3), with a valley at (3/2 3/2 -3 $\le$ L $\le$ 3). These peaks and valleys are consistent with a magnetic field-induced transfer of diffracted intensity from incommensurate parallel spin stripe ordering wavevectors, to commensurate wavevectors, indicating a polarization of the static spins participating in the parallel spin stripe structure on application of H // c.}
\label{hyspec} 
\end{figure}

\begin{figure}[tbp]
\includegraphics[width=3in]{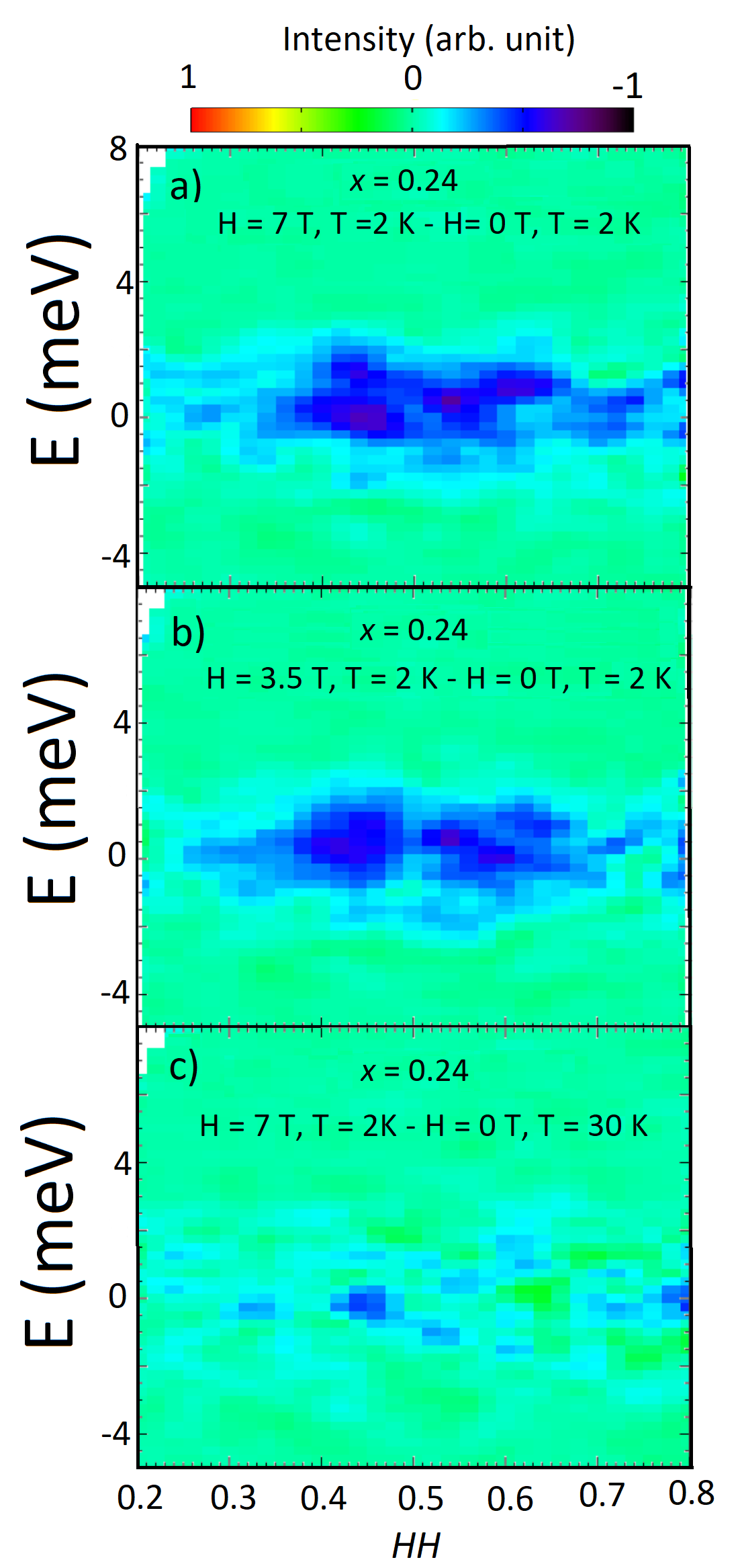}
\caption{Panels a), b) and c) show different magnetic-field difference data sets for Nd-LSCO with $x$=0.24, plotted in the energy-$HH$ reciprocal space plane.  These difference plots, made up from time-of-flight data sets, have been integrated in $L$ from -3 to 3 and in $H\bar{H}$ from -0.1 to 0.1.  Panel a) shows H = 7 T - H = 0 T, at T = 2 K, while panel b) shows 3.5 T - 0 T, at T = 2K. The net negative intensity around $HH$ = 0.4 and 0.6 in a) and b) show the suppression of the quasi-Bragg peaks due to parallel spin stripe order in an applied magnetic field H // c = 7 T and 3.5 T, respectively.  Panel c) shows the difference between data sets at H = 7 T at T = 2 K and H = 0 T at T = 30 K $>$ 2D T$_N$.  Collectively, these data sets confirm that the suppression of the incommensurate parallel spin stripe quasi-Bragg peaks is complete at H // c $>$ 2.5 T, as also concluded from our TAS data shown in Fig. \ref{fieldscan}.}
\label{hys2}
\end{figure}

Complementary time-of-flight neutron scattering measurements were also carried out on the same single crystal of Nd-LSCO with $x$ = 0.24 using the HYSPEC instrument at the Spallation Neutron Source.  Figure \ref{hyspec} b) shows an elastic scattering map in the ($H$,$K$,-3 $\le L \le$ 3) plane of reciprocal space.  This shows the difference between a data set at T = 1.5 K and H // c = 7 T and a data set at T = 1.5 K and H = 0 T.  Clear negative elastic net intensity can be observed around the (1/2, 1/2, -3 $\le L \le$ 3) and (3/2, 3/2, -3 $\le L \le$ 3) positions.  As the time-of-flight measurements also inform on the energy dependence of the scattering, we can plot a related difference plot using this data.  Figure \ref{hys2} a) shows an H // c = 7 T data set with an H = 0 T data set subtracted from it, both at T = 1.5 K.  This difference data set is now integrated in -3 $\le L \le$ 3 and -0.1 $\le H\bar{H}$ $\le$ 0.1, and it is plotted as a function of energy vs the $HH$ direction in reciprocal space in Fig. \ref{hys2} a).  Figure \ref{hys2} b) shows the same integrations, but the difference data is the subtraction of an H = 0 T data set from an H // c = 3.5 T data set, both at T = 1.5 K.  Fig. \ref{hys2} c) shows the subtraction of an H = 0 T data set for T = 30 K $>$ 2D T$_N$ from an H // c = 7 T data set at T = 1.5 K.

.

Together, the HYSPEC data sets shown in Fig. \ref{hyspec} and \ref{hys2} demonstrate that the destruction of the incommensurate elastic scattering from the parallel spin stripes is complete by H // c = 3.5 T, consistent with the TAS data which showed complete suppression by $\sim$ 2.5 T.  The effect of the magnetic field // c at H=7 T is as effective as raising the temperature to $>$ 2 $\times$ 2D T$_N$, as Fig. \ref{hys2} c) demonstrates. 

However, the elastic scattering survey of the $x$ = 0.24 single crystal in its ($H$,$K$,-3 $\le L \le$ 3) plane shows more than the destruction of the static incommensurate parallel spin stripe scattering for H // c beyond $\sim$ 2.5 T.  The reciprocal space map in Fig. \ref{hyspec} b) and cuts along (HH0) in Figs \ref{hyspec} a) and c) show a concomitant build up of scattering around commensurate wavevectors, such as (0,1,-3 $\le L \le$ 3), (1, 1, -3 $\le L \le$ 3) and (1, 2, -3 $\le L \le$ 3).  This is interpreted as the static moment participating in the parallel spin stripe structure not being destroyed, but rather being polarized by the magnetic field.

\section{The measurement of the spin gap in $La_{1.6-x}Nd_{0.4}Sr_{0.24}CuO_4$ and its correlation with T$_C$ compared to other cuprate superconductors}

Inelastic neutron scattering measurements on Nd-LSCO with $x$ = 0.24 were also carried out using TAS, at the incommensurate parallel spin stripe ordering wavevector (0.5, 0.64, 0) for energies less than 11 meV.  These measurements were carried out at base temperature, T = 1.5 K in both H = 0 T and H = 8 T // c.  Approximating the H = 8 T state as the normal state, we then expect the difference between the inelastic scattering measured in 8 T and 0 T to identify the spin gap, as spin fluctuations less than a characteristic spin gap energy are expected to be suppressed in the superconducting states, compared to the normal state.

\begin{figure}[tbp]
\linespread{1}
\par
\hspace*{-0.1in}
\includegraphics[width=3.2in]{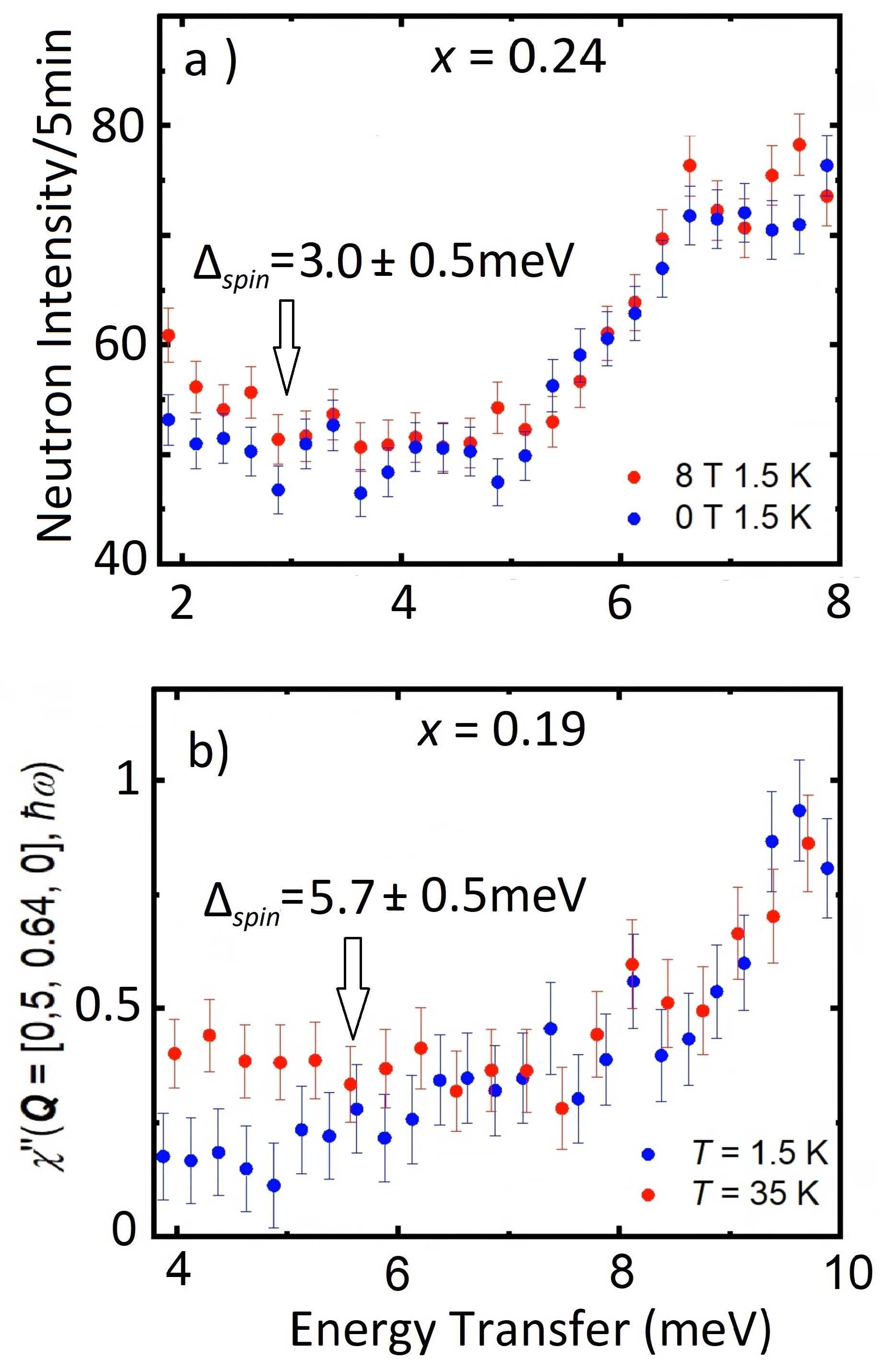}
\par
\caption{a) Inelastic neutron scattering from Nd-LSCO with $x$ = 0.24 for both H // c = 8 T and T = 1.5 K and H = 0 and T=1.5 K.  This TAS data is a constant {\bf Q} scan at the (0.5, 0.64, 0) parallel spin stripe ordering wavevector. The spin gap, $\Delta_{spin}$ = 3 $\pm$ 0.5 meV, is identified as the energy at which inelastic scattering in the (near) normal state (H // c = 8 T) no longer exceeds that in the superconducting state (H = 0). b) $\chi\prime\prime({\bf Q}, \hbar\omega)$ extracted from inelastic neutron scattering i, as reported in \cite{Kyle} is shown in zero magnetic both above, T = 35 K, and below, T = 1.5 K, superconducting T$_C$.  $\Delta_{spin}$= 6 $\pm$ 0.5 meV, is identified as the energy at which inelastic scattering in the normal state (T = 35 K) no longer exceeds that in the superconducting state (T = 1.5 K).}
\label{gapneutron} 
\end{figure}

These inelastic scattering data, in the 2 meV $\le$ $\hbar\omega$ $\le$ 8 meV regime, at H = 0 T and H = 8 T, both at T = 1.5 K, are shown in Fig. \ref{gapneutron} a). As can be seen, the inelastic scattering is systematically higher in the H = 8 T data set below 3 meV $\pm$ 0.5 meV, and this is identified as the measured spin gap, $\Delta_{spin}$ for Nd-LSCO $x$ = 0.24.  

A similar protocol has been followed to identify $\Delta_{spin}$ in other cuprate superconductors. However, as these other cuprates have superconducting T$_C$'s larger, and often much larger, than those of the Nd-LSCO family, the corresponding critical magnetic fields are also larger.  Therefore the normal state spin fluctuation spectrum is measured for T $>$ T$_C$, and the corresponding $\chi \prime\prime ({\bf Q}, \hbar\omega) = S({\bf Q}, \hbar\omega) \times (1-exp({-\hbar\omega \over kT}))$ is compared with the $\chi \prime\prime ({\bf Q}, \hbar\omega)$ at low temperatures, well below T$_C$, in order to identify $\Delta_{spin}$.  Recent work by Li \textit{et al}\cite{LiLSCO} nicely illustrates this protocol for the $x$ =0.17 and 0.21 members of the LSCO family.  This same work also compiles results for $\Delta_{spin}$ and T$_C$ across five families of cuprate superconductors, with T$_C$'s ranging from 25 K to 95 K.  

\begin {figure}[tb]
\linespread{1}
\par
\hspace*{-0.2in}\vspace*{-0.2in}\includegraphics[width=3.5in]{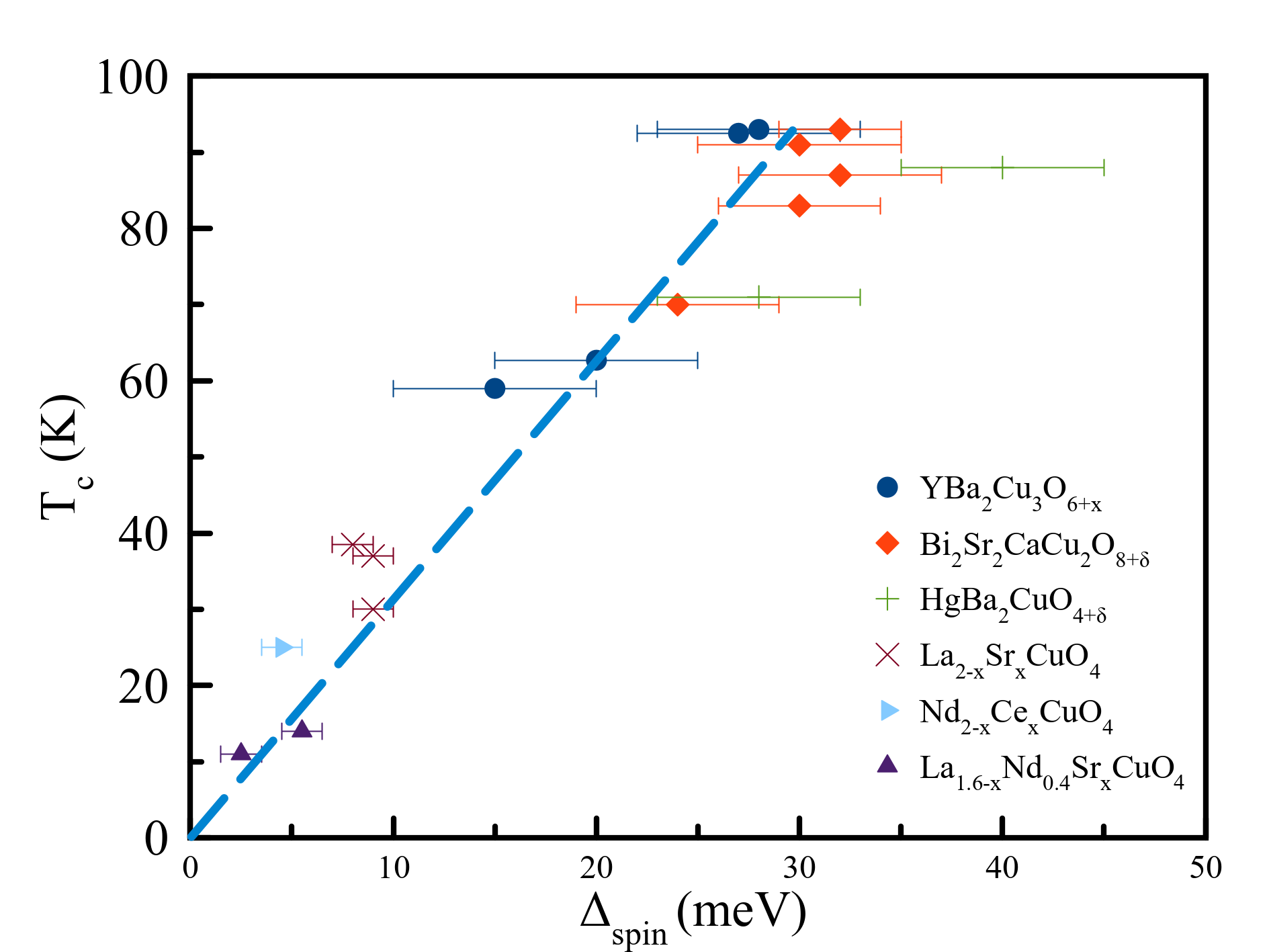}
\par
\caption{A compilation of spin gap $\Delta_{spin}$ vs sueprconducting T$_C$ is shown for five different families of cuprate superconductors.  The data points for Nd-LSCO are for the $\Delta_{spin}$ results for $x$=0.24 and 0.19 reported in this paper.  The other data points were compiled in Li et al\cite{LiLSCO}.  The overall data set is well described by a linear relationship between the two, $\Delta_{spin}$=3.5 k$_B$T$_C$, similar to that describing the energy gap vs T$_C$ relation in BCS superconductors.}
\label{allgap}
\end{figure}

We can follow this same temperature protocol to estimate $\Delta_{spin}$ for Nd-LSCO $x$ = 0.19, as data for $\chi \prime\prime ({\bf Q}, \hbar\omega)$ at T = 1.5 K and 35 K have previously been published \cite{Kyle}.  We plot this $\chi \prime\prime ({\bf Q}, \hbar\omega)$ data for Nd-LSCO $x$ = 0.19 in Fig. \ref{gapneutron} b) in the 4 meV $\le$ $\hbar$ $\omega$ $\le$ 10 meV regime.  As in Fig. \ref{gapneutron} a) for $x$ = 0.24, this is a 6 meV energy range for which the low energy spectral weight is approximately energy-independent, making it relatively easy to identify a low energy suppression of the magnetic spectral weight in the superconducting state.   We identify the dip in $\chi \prime\prime ({\bf Q}, \hbar\omega, T=1.5 K)$ below $\chi \prime\prime ({\bf Q}, \hbar\omega, T=35 K)$ at 5.7 $\pm$ 0.5 meV as $\Delta_{spin}$.  With estimates for $\Delta_{spin}$ in hand for Nd-LSCO for $x$ = 0.24 and 0.19, we can extend Li \textit{et al}'s \cite{LiLSCO} compilation of $\Delta_{spin}$ vs T$_C$ data for cuprate families down towards T$_C$=0.  These data are plotted in Fig. \ref{allgap}, along with a straight line guide-to-the-eye, which passes through $\Delta_{spin}$ = 0, T$_C$ = 0.  One can see that a good linear relation between $\Delta_{spin}$ and T$_C$, $\Delta_{spin}$ = 3.5 k$_B$ T$_C$, is obeyed across almost an order of magnitude in T$_C$. 

\section{Discussion}

The main result of this paper is that the static parallel spin stripe structure observed in Nd-LSCO with $x$ = 0.24 below 2D T$_N$=13 $\pm$ 1 K in H = 0 T, is eliminated in H // c $\ge$ 2.5 T.  On the one hand, this is a surprising result, as previous in-field diffraction studies of parallel spin stripe order have shown little or no field dependence for accessible magnetic field strengths ($\le$ 10 T).  The previous study most similar to that reported here is the study on Nd-LSCO with $x$ = 0.15 \cite{wakimotomag2003}, underdoped but approaching optimal doping. This $x$=0.15 study showed the elimination of the strong upturn of the parallel spin stripe order parameter below $\sim$ 5 K by application of a magnetic field // c.  This is interpreted as the polarization of the Nd$^{3+}$ magnetism and the removal of its influence on the low temperature spin stripe structure.  At higher temperatures in H // c = 7 T, the $x$ = 0.15 order parameter is about 20 $\%$ lower than that in H = 0, but its 2D T$_N$ remains $\sim$ 40 K.  Thus the higher temperature component of the order parameter, presumably due to Cu$^{2+}$ magnetism alone, is only modestly diminished in a H // c = 7 T field.  

Our current results on $x$ = 0.24 extend this trend significantly.  Now the low temperature upturn in the parallel spin stripe order parameter is suppressed in a 1st order fashion as a function of field, with hysteresis, as Fig. \ref{fieldscan} shows, and this initial suppression is complete by H // c $\sim$ 0.7 T.  However the complete elimination of the incommensurate parallel spin stripe quasi-Bragg peaks requires H$_C$ // c $\sim$ 2.5 T.  Thus the static parallel spin stripe order in Nd-LSCO with $x$ = 0.24 is similar to that observed in $x$ = 0.15 and 0.125, in that it is characterized by almost the same incommensurate parallel spin stripe wavevector, but it is much more fragile, with 2D T$_N$ a factor of 3 to 4 lower, and an H$_C$ // c sufficiently reduced from $x$ = 0.15 and 0.125, such that this effect is now observable with modest magnetic field capabilities.  

It is interesting to note that while our elastic neutron scattering measurements of parallel spin stripes in zero magnetic field do not obviously reflect a pseudogap QCP, elastic neutron scattering in low but finite magnetic field may show exactly that - the disappearance of magnetic-field-robust parallel spin stripe order near $x^*$ = 0.23.

An important corollary to the observed field-dependence of the static parallel spin stripe order in Nd-LSCO $x$ = 0.24 is that in-field transport and thermodynamic measurements for samples beyond optimal doping, such as those reported in \cite{michon2019thermodynamic} and \cite{fang2020fermi}, are not expected to see spin stripe order as the same magnetic fields required to overcome superconductivity and establish the normal state will destroy the parallel spin stripe order.  As shown in Fig. \ref{hyspec}, the static moments participating in the spin stripes structure are not destroyed, but rather appear to be polarized in modest H // c. 

Finally, we took advantage of relatively low critical field // c for superconductivity in order to examine the low energy spin fluctuations at the parallel spin stripe ordering wavevector in both a superconducting ground state (H = 0) and in a state with most of its volume corresponding to the normal state (H = 8 T).  This is a novel protocol for identifying the spin gap, $\Delta_{spin}$, compared with the more typical protocol of looking at the spin fluctuations above and below T$_C$ in zero field.  However, the use of the field-induced normal state has the advantage that it allows all the spin fluctuation measurements to be performed at base temperature, and hence no correction for the Bose factor, which relates $S({\bf Q}, \hbar\omega)$ to $\chi\prime\prime({\bf Q}, \hbar\omega)$, is required.  The Bose factor correction is not difficult to apply, but it must be applied to $S({\bf Q}, \hbar\omega)$ alone, and thus $S({\bf Q}, \hbar\omega)$ must be clearly differentiated from background scattering in order to use the temperature-difference protocol.

We have used the field protocol to estimate $\Delta_{spin}$ in the $x$ = 0.24 sample, and the temperature-difference protocol to do the same for $x$ = 0.19, using previously published data \cite{Kyle}. This allows us to add two data points at the low field extreme of the $\Delta_{spin}$ vs T$_C$ compilation initiated by Li et al\cite{LiLSCO}.  We find that these estimates fit with the linear trend between these two observables in five different families of cuprate superconductors, covering almost a decade in T$_C$.  Remarkably this linear relation is given by $\Delta_{spin}$ = 3.5 k$_B$ T$_C$, implying a role for $\Delta_{spin}$ akin to 2 $\times \Delta$ as measured in tunneling experiments on BCS superconductors. 

\section{Conclusions}

Elastic neutron scattering measurements on single crystal Nd-LSCO with $x$ = 0.24 in the presence of a magnetic field, H // c, show that the previously observed parallel spin stripe incommensurate quasi-Bragg peaks are reduced to zero intensity for H $\ge$ 2.5 T.  Their low temperature enhancement below T $\sim$ 5 K in H = 0, is eliminated in a 1st order fashion with H for H // c $\le$ 0.7 T.  While the effect of the field on the low temperature enhancement of the parallel spin stripe order parameter is consistent with earlier work at $x$ = 0.15 \cite{wakimotomag2003}, the complete elimination of the order parameter for fields above $\sim$ 2.5 T is surprising given the lack of field dependence previously observed near $x$ = 0.125 \cite{changmagnetic2008}, and the enhancements of the order parameter in field observed in underdoped LSCO\cite{lscoenhance}.  We conclude that while the incommensurate parallel spin stripe structure in Nd-LSCO with $x$ = 0.24 is similar to that exhibited in $x$ = 0.125 and 0.15, it is much more easily perturbed by either temperature or magnetic field.  This work has implications for transport and thermodynamic measurements made in the presence of large magnetic fields, so as to establish normal state properties at low temperatures in Nd-LSCO at relatively high hole-doping.

Low energy inelastic neutron scattering measurements at the parallel spin stripe ordering wavevevector in Nd-LSCO $x$ = 0.24 in H = 8 T allow us to estimate a spin gap, $\Delta_{spin}$=3.0 $\pm$ 0.5 meV.  We combine this result, with an analysis on preexisting Nd-LSCO $x$ = 0.19 data, to fill out the low T$_C$ region of a $\Delta_{spin}$ vs T$_C$ compilation across five families of cuprate superconductors, and observe a linear relationship between them.  Intriguingly, this relationship is roughly given by $\Delta_{spin}$ = 3.5 k$_B$T$_C$.

\section{Acknowledgments}

We thank Amirreza Ataei, Takashi Imai, Graeme Luke,Young June Kim, Steven Kivelson and Louis Taillefer for useful and stimulating discussions. This work was supported by the Natural Sciences and Engineering Research Council of Canada. This research used resources at High Flux Isotope Reactor and the Spallation Neutron Source, DOE Office of Science User Facilities operated by the Oak Ridge National Laboratory (ORNL).

%

\end{document}